\documentclass[runningheads]{llncs}
\usepackage[T1]{fontenc}
\usepackage{graphicx}
\usepackage{hyperref}
\usepackage{subcaption}
\usepackage{xcolor}
\definecolor{keywordColor}{rgb}{0.0, 0.0, 0.6}
\definecolor{commentColor}{rgb}{0.4, 0.4, 0.4}
\definecolor{stringColor}{rgb}{0.0, 0.5, 0.0}
\definecolor{variableColor}{rgb}{0.5, 0.0, 0.5}
\definecolor{uriColor}{rgb}{0.0, 0.4, 0.8}
\definecolor{pastelblue}{rgb}{0.6, 0.8, 1.0}
\definecolor{lightgray}{gray}{0.85}
\usepackage{caption}
\usepackage{listings}
\usepackage{float}
\lstdefinelanguage{sparql}
{
  basicstyle=\fontencoding{T1}\small\ttfamily\color{black}, 
  sensitive=false, 
  morekeywords={SELECT,CONSTRUCT,DESCRIBE,ASK,WHERE,FROM,NAMED,PREFIX,BASE,OPTIONAL,FILTER,GRAPH,LIMIT,OFFSET,SERVICE,UNION,EXISTS,NOT,BINDINGS, BIND,COUNT,MINUS,a, REGISTER, WINDOW, ON, RSTREAM,ISTREAM,DSTREAM, RANGE, STEP, ORDER, STREAM, GROUP, BY, AS, SEQ, FIRST, LAST, WIN, EVENT, LND, MATCH},
  keywordstyle=\color{keywordColor},
  alsoletter=\#-,
  morecomment=[l]{\#\ }, 
  commentstyle=\color{commentColor},
  morestring=[b]", 
  stringstyle=\color{stringColor},
  moredelim=*[s][\color{variableColor}]{?}{\ },
  moredelim=*[s][\color{uriColor}]{<http://}{>},
  moredelim=*[s][\color{uriColor}]{<https://}{>},
  moredelim=*[s][\color{black}]{\ P}{\ },
  moredelim=*[s][\color{black}]{NOW-P}{\ },
  moredelim=*[s][\color{black}]{&&}{\ },
  captionpos=b, 
  numberstyle=\tiny\ttfamily, 
  frame=trbl, 
  showstringspaces=false, 
  breaklines=true, 
  extendedchars=true, 
  tabsize=2, 
  columns=fixed, 
  keepspaces=true, 
  aboveskip=15pt, 
  belowskip=0pt,
  escapechar=|
}

\lstdefinelanguage{mapping}
{
  basicstyle=\fontencoding{T1}\footnotesize\ttfamily\color{black}, 
  sensitive=false, 
  morekeywords={mappingId, target, source},
  keywordstyle=\color{keywordColor},
  alsoletter=\#-,
  morecomment=[l]{\#\ }, 
  commentstyle=\color{commentColor},
  morestring=[b]", 
  stringstyle=\color{stringColor},
  moredelim=*[s][\color{variableColor}]{\{}{\}},
  moredelim=*[s][\color{uriColor}]{<http://}{>},
  moredelim=*[s][\color{uriColor}]{<https://}{>},
  moredelim=*[s][\color{black}]{\ P}{\ },
  moredelim=*[s][\color{black}]{NOW-P}{\ },
  moredelim=*[s][\color{black}]{&&}{\ },
  captionpos=b, 
  numberstyle=\tiny\ttfamily, 
  frame=trbl, 
  showstringspaces=false, 
  breaklines=true, 
  extendedchars=true, 
  tabsize=2, 
  columns=fixed, 
  keepspaces=true, 
  aboveskip=15pt, 
  belowskip=0pt,
  escapechar=|
}

\lstdefinelanguage{python}
{
  basicstyle=\fontencoding{T1}\small\ttfamily\color{black}, 
  sensitive=false, 
  morekeywords={},
  keywordstyle=\color{keywordColor},
  alsoletter=\#-,
  morecomment=[l]{\#\ }, 
  commentstyle=\color{commentColor},
  morestring=[b]",
  morestring=[b]',
  stringstyle=\color{stringColor},
  moredelim=*[s][\color{variableColor}]{?}{\ },
  moredelim=*[s][\color{uriColor}]{<http://}{>},
  moredelim=*[s][\color{uriColor}]{<https://}{>},
  moredelim=*[s][\color{black}]{\ P}{\ },
  moredelim=*[s][\color{black}]{NOW-P}{\ },
  moredelim=*[s][\color{black}]{&&}{\ },
  captionpos=b, 
  numberstyle=\tiny\ttfamily, 
  frame=trbl, 
  showstringspaces=false, 
  breaklines=true, 
  extendedchars=true, 
  tabsize=2, 
  columns=fixed, 
  keepspaces=true, 
  aboveskip=15pt, 
  belowskip=0pt,
  escapechar=|
}

\lstdefinestyle{sparqlStyle}{
 numberstyle=\tiny\color{lightgray},
 morecomment=[l][\color{pastelblue}]{\%},
 breakatwhitespace=false,         
 breaklines=true,  
 alsoletter={\#},               
 keepspaces=false,                 
 numbersep=5pt,                  
 showspaces=false,                
 showstringspaces=false,
 showtabs=false,                  
 mathescape=true,
 mathescape=true,
 columns=fixed,
 basewidth=0.5em,
 escapeinside={(*}{*)},
 showstringspaces=false,
 tabsize=2,
 breaklines=true,
 frame=tbrl,
 frameround=ffff,
 captionpos=b,
 xleftmargin=0em,
 framexleftmargin=0em,
 numbers=none,
 numberstyle=\tiny\ttfamily,
 escapechar=|
}

\floatstyle{plain}  
\newfloat{listing}{t}{lop}
\floatname{listing}{Listing}
\usepackage{booktabs}
\usepackage{listings}
\usepackage{multirow}
\usepackage{tabularx}

\begin{document}
\title{LLM-Enhanced Semantic Data Integration of Electronic Component Qualifications in the Aerospace Domain}

%

\titlerunning{LLM-Enhanced Semantic Integration of Heterogeneous Qualification Data}

\author{Antonio De Santis\inst{1}\and
Marco Balduini\inst{2,3}\and
Matteo Belcao\inst{2}\and
Andrea Proia\inst{4} \and \\
Marco Brambilla\inst{1} \and
Emanuele Della Valle\inst{1}
}

\authorrunning{A. De Santis et al.}
%
\institute{Politecnico di Milano, DEIB, I-20133 Milano, Italy \\
\email{\{antonio.desantis,marco.brambilla,emanuele.dellavalle\}@polimi.it} \and 
Quantia Consulting, Milano, Italy \\
\email{\{marco.balduini,matteo.belcao\}@quantiaconsulting.com}\and 
Motus ml, Milano, Italy \and
Thales Alenia Space, Roma, Italy \\
\email{andrea.proia@thalesaleniaspace.com}}
\maketitle              
\begin{abstract}
Large manufacturing companies face challenges in information retrieval due to data silos maintained by different departments, leading to inconsistencies and misalignment across databases. This paper presents an experience in integrating and retrieving qualification data for electronic components used in satellite board design. Due to data silos, designers cannot immediately determine the qualification status of individual components. However, this process is critical during the planning phase, when assembly drawings are issued before production, to optimize new qualifications and avoid redundant efforts.
To address this, we propose a pipeline that uses Virtual Knowledge Graphs for a unified view over heterogeneous data sources and LLMs to enhance retrieval and reduce manual effort in data cleansing. The retrieval of qualifications is then performed through an Ontology-based Data Access approach for structured queries and a vector search mechanism for retrieving qualifications based on similar textual properties. We perform a comparative cost-benefit analysis, demonstrating that the proposed pipeline also outperforms approaches relying solely on LLMs, such as Retrieval-Augmented Generation (RAG), in terms of long-term efficiency.

\keywords{Knowledge Graphs \and Large Language Models \and Data Integration \and Space Industry}
\end{abstract}
\section{Introduction}
Dealing with data distributed across silos is a persistent challenge for large manufacturing companies, where different departments maintain systems that evolve independently over time. This fragmentation leads to inconsistencies and heterogeneity, making information retrieval a challenging task. This is particularly pronounced in industries like aerospace, where products are highly complex and are produced in low volumes. As a result, most of the data is manually recorded in spreadsheet-like databases, which are prone to errors and heterogeneity.

The experience presented in this article focuses on the problem of qualification of components, i.e., ensuring that a produced item consistently meets the expected quality and reliability standards for its use in different scenarios. 
More precisely, we consider the qualification of electronic components (e.g., capacitors, resistors) used as part of electronic boards that are installed on Thales Alenia Space's satellites. Given a certain component, determining whether it is qualified according to space agencies regulations and the type of qualification is challenging due to the need to cross-reference two distinct and heterogeneous data sources. However, accessing past qualifications is essential for optimizing qualification planning and preventing unnecessary redundancies.
To address this use case, we adopted a hybrid approach that enables semantic data integration and access across heterogeneous sources by combining Semantic Web technologies and Large Language Models (LLMs).
Our solution leverages Virtual Knowledge Graphs (VKGs) to provide a unified semantic view over heterogeneous sources while allowing symbolic queries through SPARQL and utilizes LLMs to enhance data cleansing, handle unstructured textual content, and support similarity-based retrieval via vector search. 

\medskip\noindent\textbf{Structure of the Work.}
The rest of this paper is structured as follows.
Section~\ref{related} presents a review of related work. Section~\ref{mot} describes our motivating use case. Section~\ref{methodology} explains our proposed methodology, while a comparison with a RAG-based approach and an analysis of the solution's potential impact are provided in Section~\ref{impl}. Section~\ref{lessons} discusses the deployment experience and the lessons learned, while Section~\ref{concl} concludes the paper with directions for future work.

\section{Related Work}
\label{related}
\noindent\textbf{Ontology-based Data Integration.}
Virtual Knowledge Graphs (VKGs)~\cite{vkg} have already been utilized for data integration in various industrial settings, as they allow unified access to different data sources via a virtual semantic layer (i.e., an ontology).
For instance, in the aerospace domain, VKGs have been applied to integrate and query test reports data of Printed Circuit Boards (PCBs) at Thales Alenia Space~\cite{tasi}. 
Statoil has also adopted a similar approach utilizing the \emph{Ontop}~\cite{ontop} framework to integrate diverse data sources \cite{statoil}, significantly benefiting geologists involved in oil and gas exploration and production.
VKGs and Ontop have also been used to integrate Bosch manufacturing data for analyzing the production pipeline of electronic components~\cite{bosch_integration}.
Similarly, Ontop has been used for a global manufacturing company, enabling semantic integration of heterogeneous data sources, facilitating tool management and energy consumption monitoring~\cite{manufacturing}. In this paper, we expand on this body of applied research by including LLMs in the integration pipeline and adapting these technologies to solve a concrete industrial challenge.

\medskip\noindent\textbf{Knowledge Engineering using LLMs.}
In recent years, Large Language Models (LLMs) have advanced significantly, particularly with the introduction of models such as OpenAI's GPTs~\cite{gpt3,gpt4}. LLMs have found applications in numerous knowledge engineering tasks, such as knowledge extraction~\cite{tasi,extraction}, ontology engineering~\cite{Crum2024,SHIMIZU2025100862,fathallah2024neon,zhang2024ontochat}, and information retrieval~\cite{rag_openbook,edge2025localglobalgraphrag}. Regarding applied research in real-world scenarios, LLMs have been increasingly integrated into domain-specific knowledge engineering workflows as they enhance the accessibility and usability of heterogeneous data and achieve automation and efficiency while incorporating human oversight to ensure precision. For instance, in the healthcare and life sciences domain, LLMs have been leveraged for both ontology learning~\cite{parkinson,fathallah2024llms4life} and knowledge graph population~\cite{spires}. Similarly, in the scholarly domain, LLMs have been employed to semi-automatically populate and validate metadata for academic conferences in Wikidata~\cite{scholarly_wikidata} as well as for scholarly ontology generation~\cite{aggarwal2024largelanguagemodelsscholarly}. LLMs have also played a role in the cultural heritage domain, automatically populating a cultural tourism knowledge graph with human oversight~\cite{cupe}.
Regarding industrial settings, LLMs have already been used at Thales Alenia Space for automating extraction and validation of textual test reports of PCBs, achieving significant time-saving~\cite{tasi}. Siemens has also explored LLM-guided ontology term definition generation, where LLMs assist in formulating precise definitions for domain-specific terms~\cite{siemensllm}. Additionally, automated ontology generation has been used to enable zero-shot defect identification in manufacturing by leveraging structured domain knowledge~\cite{defect}.

\section{Motivating Use Case}
\label{mot}
In this section, we explore the motivations behind our case study. We then describe the structure and formats of Thales Alenia Space's qualification data and the different types of qualifications we are aiming to retrieve.

\medskip\noindent\textbf{Context and Motivation.}
In the aerospace sector, mission success and safety depend heavily on the quality and compliance of electronic assemblies, such as capacitors, resistors, and integrated circuits, used in satellites and other spacecraft. Each of these components must adhere strictly to rigorous qualification standards defined by aerospace regulatory frameworks established by entities such as the European Space Agency (ESA) and the European Cooperation for Space Standardization (ECSS) \cite{ecss-q-st-70-61c}.

Furthermore, the qualification of electronic components involves extensive and resource-intensive verification processes, including mechanical, thermal, and electrical tests, to ensure that each component meets the demanding conditions of space environments. Given these high costs associated with qualification processes, it is of strategic importance to try to reuse existing qualifications whenever possible.
Consequently, access to previous qualifications is essential during the planning phase, when assembly drawings are issued before production, to optimize the planning of new qualifications and avoid the proliferation of new ones while ensuring compliance with stringent aerospace standards.

\medskip\noindent\textbf{Qualification Data.}
The main challenge addressed in this paper is that the retrieval of qualification information within the company requires cross-referencing two different internal databases:  

\begin{itemize}
  \item A \textit{Product Lifecycle Management Database (PLM-DB)}, which comprehensively lists all electronic components utilized by Thales Alenia Space in the production of PCBs, along with their technical characteristics.
  \item A \textit{Qualification Catalog (QC)}, which tracks electronic components that have either completed or are undergoing a qualification process. The status of the qualification can be \textit{Closed}, \textit{Ongoing}, \textit{Failed}, or \textit{Obsolete}.
\end{itemize}

Both databases, but QC in particular, are manually maintained and updated by designers and technologists, resulting in numerous inconsistencies, heterogeneities, and potential data-entry errors. Furthermore, essential information is often hidden within free-text columns, making the task of accurately cross-referencing components between PLM-DB and QC highly complex and time-consuming. Before the deployment of the solution described in this paper, Thales Alenia Space designers managed this challenge by manually searching through web forms of both databases, significantly increasing the risk of oversight and inefficiency given that PLM-DB contains tens of thousands of entries while QC's size is in the thousands. Some examples of entries from the two databases, featuring a small subset of columns, are provided in Table~\ref{tab:example}. 

In the PLM-DB, the combination of “Part Number”, “Package Code”, “Subpackage Code”, and “Manufacturer” identifies a certain entry, while other columns contain various characteristics such as “Family”, “Pitch”, “Dimension”, and many others. On the other hand, for QC, an entry is identified by a “Qualification Number”, while other columns contain a subset of component characteristics, the status of the qualification, and a column “Notes” which contains other information in free-text format. In the vast majority of cases, the “Part Number (PN)” is buried within a textual note in this column, as there is no dedicated column for PN in QC. An example of text contained in “Notes” is the following: \textit{“R1 (pn P3333333) double component soldered on double pad, soldered in HS (Hot Soldering) with a stand-off of 0.2–0.3 mm on polyimide”}. Another example of heterogeneity is the inconsistent representation of manufacturer names, which might appear as “ABC”, “ABC Corp”, “ABC Inc.”, or other variations.

\begin{table}
    \centering
    \caption{Examples of records from the PLM-DB and QC.}
    \label{tab:example}
    \bigskip

    \textit{Product Lifecycle Management Database (PLM-DB)} \\
    \vspace{0.25em}
    \begin{tabular}{l|l|l|l|l|l}
        \toprule
        Part Number (PN) & Package Code & Subpackage Code & Manufacturer & Family & Pitch \\
        \midrule
        P1111111 & FP1  & a1 & ABC & Hybrid & 1.27 \\
        P2222222 & C2  & x2 & XYZ & Capacitor & 2.2 \\
        P3333333 & R1  & a3 & ABC & Resistor & 1.92 \\
        \bottomrule
    \end{tabular}
    
    \vspace{1em}

    \textit{Qualification Catalog (QC)} \\
    \vspace{0.25em}
    \begin{tabular}{l|l|l|l|l|l}
        \toprule
        Qual. Number & Package Code & Subpackage Code & Manufacturer & Qual. Status & Notes\\
        \midrule
        qc1 & FP1  & a1 & ABC Corp & Closed & ... \\
        qc2 & C2 & x2 & XYZ Inc. & Closed & ...\\
        qc3 & R1  & a3 & ABC Inter. & Ongoing & ...\\
        \bottomrule
    \end{tabular}
\end{table}

Given these structural discrepancies and different naming conventions, retrieving previous qualifications is challenging, as it requires integrating and querying these databases. Furthermore, even when structural differences are addressed through a local integration view, the data remains difficult to merge due to quality issues. For instance, the heterogeneity in the “Manufacturer” column would hinder the effective use of standard join operations.

\medskip\noindent\textbf{Qualification Types and Requirements.}
Qualifications obtained for specific components can potentially be applied to other components, provided they adhere to established aerospace standards and demonstrate comparability. This comparability includes similarities in critical attributes such as package design, terminal finish, thermal characteristics, assembly processes, and intended operating conditions. Specifically, our case study includes three types of qualifications under ECSS guidelines (Section 13.8 in \cite{ecss-q-st-70-61c}):

\begin{itemize}
\item \textit{Direct qualification}: Achieved when the attributes “Part Number”, “Package Code”, “Subpackage Code”, and “Manufacturer” in the PLM-DB correspond exactly with the respective attributes in QC, indicating that the component in question has previously undergone the full formal qualification process.
\item \textit{Qualification by similarity}: Achieved when the attributes “Package Code”, “Subpackage Code”, and “Manufacturer” in the PLM-DB precisely match those present in QC, however, the “Part Number” is different. In this scenario, the component is not identical to the originally qualified unit but exhibits sufficiently comparable physical, electrical, and manufacturing characteristics to justify extending the qualification.
\item \textit{Alternative qualification}: 
The specific criteria for alternative qualification are dependent on the component type and always require detailed technical review and validation by qualified experts\footnote{This type of qualification serves more as a suggestion rather than a definitive answer.}. 
For instance, Flat Packages (FP) require that the “Package Code” and “Pitch” match between PLM-DB and QC, that the “Pin Dimension” is within a range of ±5 microns from the original qualified component, and that the assembly process is also equivalent.
\end{itemize}

Considering this, the requirements for a qualification retrieval tool must include the capability to automatically identify direct and by-similarity qualifications, as well as to propose alternative qualification candidates, which must subsequently be reviewed and validated by a designer or technologist.

\medskip\noindent\textbf{Data Confidentiality.} The exemplary data shown in the paper is not disclosed in its original form to protect Thales Alenia Space’s confidentiality. We modified free-text descriptions, codes, and numerical values, ensuring the structure and syntax remained intact without disclosing any confidential information.

\section{Methodology}
\label{methodology}
\begin{figure*}[t]
    \centering
    \includegraphics[width=1\textwidth]{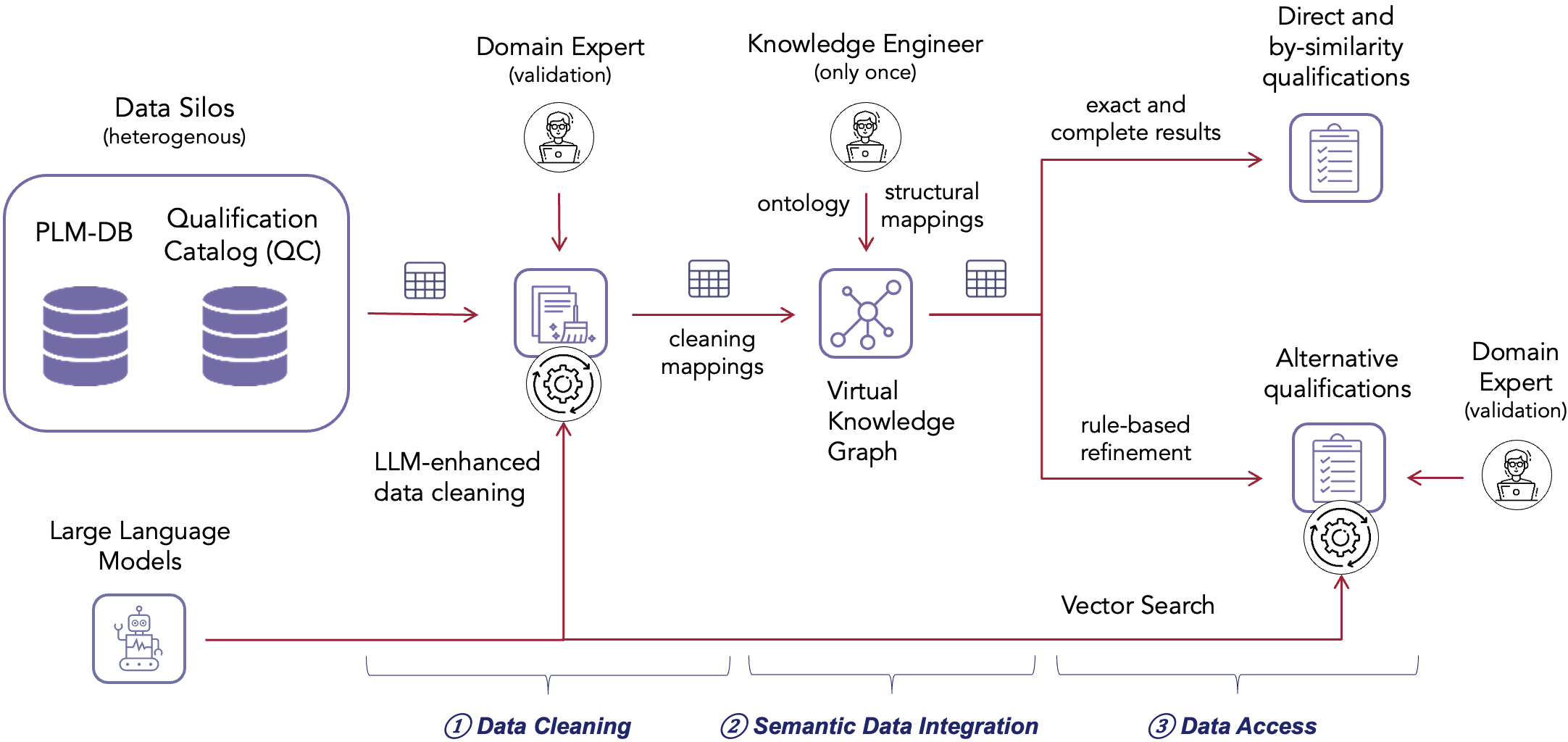}
    \caption{Overview of the proposed LLM-enhanced semantic integration pipeline. The approach integrates two heterogeneous data silos (PLM-DB and QC) by combining LLMs and VKGs. LLMs assist in semi-automated data cleaning, including normalization and extraction of key fields, while knowledge engineering is performed once to create a semantic layer over the data. Direct and by-similarity qualifications are retrieved via exact symbolic queries over the VKG, whereas alternative qualifications are retrieved through vector search and refined using rules based on the component type and finally validated by a domain expert.}
    \label{fig:pipeline}
\end{figure*}

This section describes the methodology we adopted for implementing our qualification retrieval tool. This process, which is outlined in Figure \ref{fig:pipeline}, is structured into three main phases, each executed in close collaboration with domain experts from Thales Alenia Space:
\begin{itemize}
\item \textit{Data Cleaning:} LLMs are used to streamline data cleansing by addressing heterogeneities and inconsistencies.
\item \textit{Semantic Data Integration:} A one-time knowledge engineering process is performed to define the semantic model and mappings to integrate heterogeneous data sources into a unified VKG.
\item \textit{Data Access:} The retrieval of direct and by-similarity qualifications is then achieved via SPARQL queries. Database records are also vectorized to allow retrieval of alternative qualifications via vector-based search.
\end{itemize}

\begin{figure*}[t]
    \centering
    \includegraphics[width=1\textwidth]{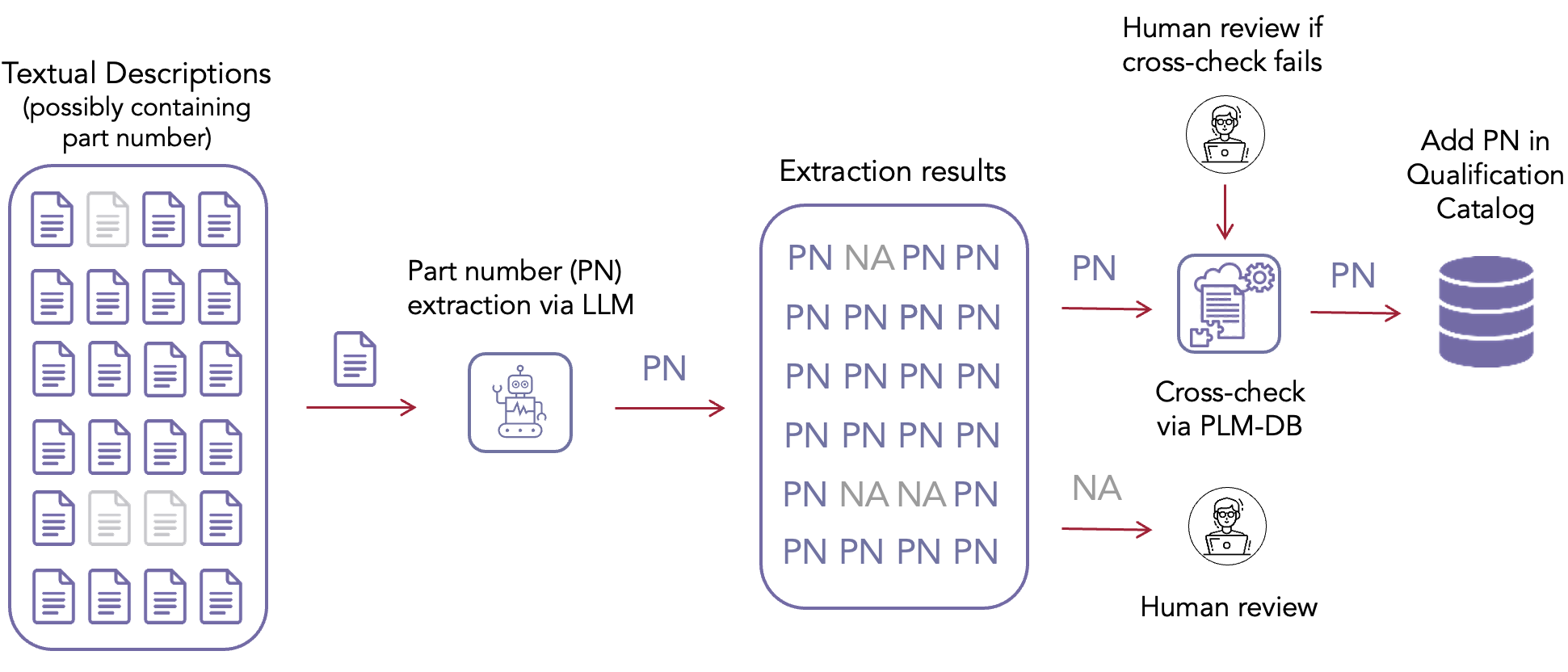}
    \caption{Overview of the pipeline for semi-automated extraction of Part Numbers (PN) from textual descriptions using LLMs. Textual entries, potentially containing Part Numbers, are first processed by an LLM that automatically extracts candidate PNs. These candidates are subsequently cross-checked against the PLM-DB to validate correctness. Entries that cannot be verified through this step are flagged and sent to human domain experts for manual review, including those for which a candidate PN could not be extracted (indicated as NA). Successfully validated PNs are then added to QC.}
    \label{fig:pn_pipeline}
\end{figure*}

\medskip\noindent\textbf{Data Cleaning. }
The first phase focuses on data cleansing to address heterogeneities and structural differences between the two databases, with particular attention to the columns required for the automatic retrieval of both direct and by-similarity qualifications. This includes standardizing column names and resolving inconsistent naming conventions in the “Manufacturer” column. Additionally, a new column for the “Part Number (PN)” must be added to QC.
While column names can be easily standardized manually, addressing the data entries within the “Part Number” and “Manufacturer” columns follows a semi-automated approach, combining LLMs with human-in-the-loop validation to ensure correctness. 
We used GPT-OSS-120B for this purpose, running on-premise to ensure data security and compliance. 

For the “Manufacturer” column, we ask the model to generate cleaning rules in a semi-automated workflow. All unique manufacturer names are first extracted from the two databases. Then, using a few-shot prompt, the LLM is asked to infer normalization rules (i.e., merging variations of the same name). A simple example of such a rule can be: \[
\{\texttt{ABC Corp},\ \texttt{ABC},\ \texttt{ABC Inc.},\ \texttt{ABC International}\} \longrightarrow \texttt{ABC}
\]

The model is also prompted to format the response as JSON so that a cross-check can be performed on-the-fly to ensure that none of the names are hallucinated and that the LLM is not ignoring any of the provided names. 
These rules are subsequently reviewed and refined by domain experts. Their feedback can also be incorporated to enrich the prompting context; for example, in cases where country-specific branches of the same manufacturer must remain distinct, this information can be included in the prompt to guide the model accordingly.

In our case study, out of 466 unique manufacturer names extracted from the dataset, GPT-OSS-120B generated 50 normalization rules. Of these, 9 (approximately 18\%) required further refinement by domain experts.  Furthermore, in a few notable instances, the model correctly identified and merged manufacturer names that were lexically dissimilar but referred to the same company, typically due to historical name changes or corporate acquisitions. Cleaning rules are stored in a separate SQL table without modifying the original manufacturer name, as it may potentially be relevant for tasks other than qualification retrieval. This table is later used to automatically generate cleaning mappings for the VKG. Despite the LLM-based cleaning being far from perfect, the semi-automated approach still proved more efficient than conducting the process entirely manually. For instance, the rule refinement process was completed in a single 2-hour workshop with company stakeholders, whereas, based on our prior experience, asking them to perform the same cleaning entirely manually would have required at least several days of work.

Considering the “Part Number”, it is not explicitly listed in QC and therefore must be extracted from unstructured textual descriptions within the “Notes” column and added as a stand-alone column. The overall process is summarized in Figure \ref{fig:pn_pipeline}. The extraction is performed automatically using GPT-OSS-120B and a few-shot prompt. Once a candidate PN is identified, it is cross-referenced with the PLM-DB to verify both its existence and the fact that it is actually the one that underwent the qualification in question. This latter check is done by comparing the “Package Code”, “Subpackage Code”, and “Manufacturer” columns. However, a human-in-the-loop still remains necessary to handle cases where the cross-check fails or the LLM fails to extract a PN from the textual description. In practice, manual intervention was mainly required because, in approximately 2\% of qualifications, the PN is not present in the description and must be looked up in the qualification documents. In a handful of additional cases, the PN was present but the model failed to comply with the requested output format, so the PN was added manually on-the-fly. Overall, validating PNs for all qualifications required only a few hours of expert time whereas performing the same extraction and verification fully manually for thousands of qualifications would have required several weeks.

\begin{figure*}[t]
    \centering
    \includegraphics[width=1\textwidth]{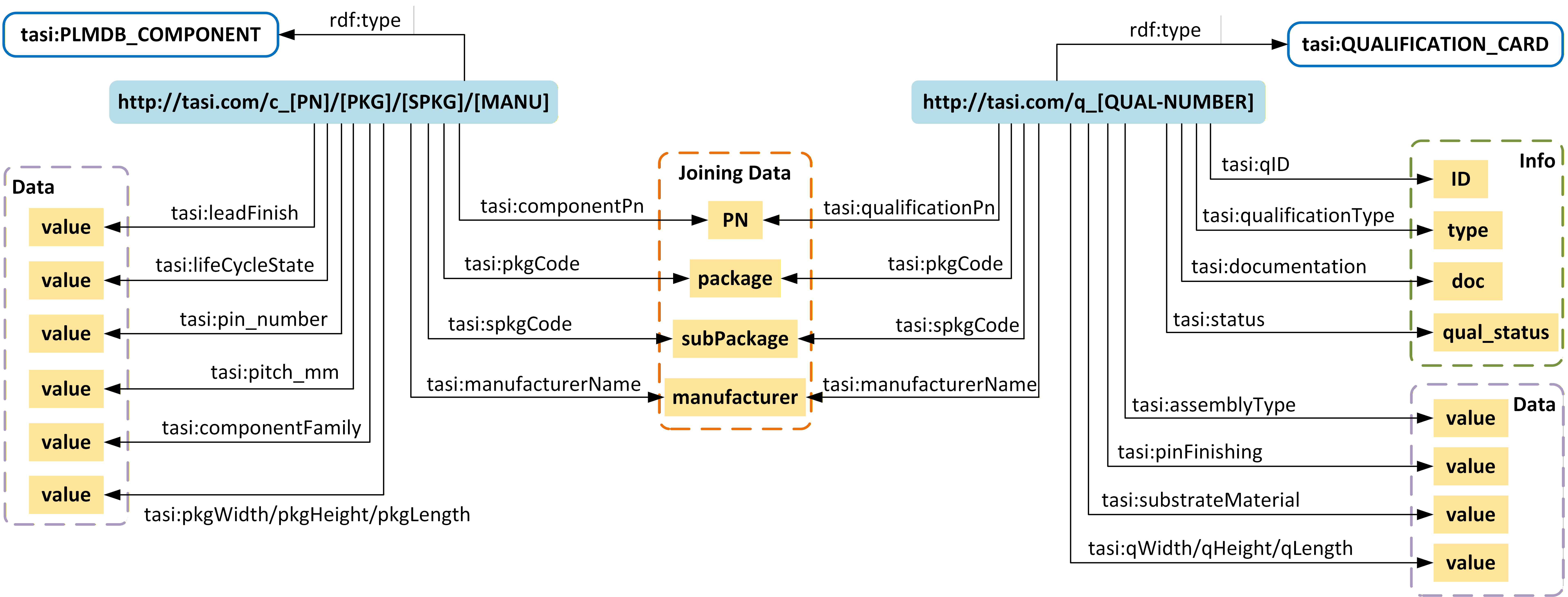}
    \caption{High-level overview of the VKG semantic model for qualification retrieval. It shows the two main entities (\texttt{PLMDB\_COMPONENT} and \texttt{QUALIFICATION\_CARD}) and the core join attributes (PN, manufacturer, package and subpackage) used for direct and by-similarity matching. Additional domain-specific properties are omitted for readability.}
    \label{fig:ontology}
\end{figure*}

\medskip\noindent\textbf{Semantic Data Integration. }
Central to the VKG paradigm is the development of an ontology to model the concepts, entities, and relationships relevant to electronic components in PLM-DB and their qualification data. This ontology serves as the foundation for constructing a VKG using the Ontop framework. Figure~\ref{fig:ontology} provides a high-level overview of the main classes and properties used.
The ontology for this use case is not particularly complex, with the two primary entities being \texttt{PLMDB\_COMPONENT} and \texttt{QUALIFICATION\_CARD}. The former represents components from PLM-DB, and exposes the core attributes required for retrieval as data properties such as \texttt{componentPn}, \texttt{manufacturerName}, \texttt{pkgCode}, and \texttt{spkgCode}, together with other physical characteristics (e.g., \texttt{leadFinish}, \texttt{rawMaterial}). Similarly, a \texttt{QUALIFICATION\_CARD} includes the corresponding identifiers (\texttt{qualificationPn}, \texttt{manufacturerName}, \texttt{pkgCode}, \texttt{spkgCode}) and qualification metadata such as \texttt{status}, \texttt{qualificationType}, and \texttt{qualificationDesc}. 

The ontology is then linked with the underlying relational databases by defining a series of mappings. To handle the heterogeneity of the manufacturer name, we define an Ontop Lens\footnote{An Ontop Lens is a relational view defined at the level of Ontop instead of at the level of the underlying database, specified upon a relational table or other lenses. Further details at: \url{https://ontop-vkg.org/guide/advanced/lenses.html}}, which performs a virtual normalization by joining the original database with the SQL table containing the LLM-generated cleaning rules. This ensures that all variants referring to the same manufacturer are mapped to a canonical name. The Ontop Lens can directly be used in a mapping as shown in Listing \ref{lst:plmdb-qualification-manufacturer-mapping}. 
The target of the mapping is a URI that identifies the qualification within the ontology, while the source is a SQL query that fetches the data from the virtual view generated by the Lens. 
When the mapping is executed, the \texttt{manufacturerName} is filled with the cleaned manufacturer name, allowing us to enable access to an RDF graph without materializing or modifying the underlying data sources. 

\renewcommand{\thelstlisting}
{\arabic{lstlisting}}
\begin{lstlisting}[language=mapping,
  caption={The mapping for the manufacturer of the \texttt{QUALIFICATION\_CARD}.},
  label={lst:plmdb-qualification-manufacturer-mapping},
  abovecaptionskip=10pt]
@prefix : <http://tasi.com/> .
@prefix tasi: <http://tasi.com#> .

mappingId Qualification_Manufacturer
target :q_{number} tasi:manufacturerName {canonical_manufacturer_name}.
source SELECT number, canonical_manufacturer_name
       FROM lenses.qualification_manufacturer
\end{lstlisting}

\medskip\noindent\textbf{Data Access. } With the VKG in place, structured access to qualification data is achieved directly through SPARQL queries over the ontology. The query is dynamically rewritten into executable SQL by Ontop. The SQL answer is then translated back to RDF or SPARQL result format, avoiding physically
materializing the PLM-DB and QC into a knowledge graph. We adopt this approach to extract both direct and by-similarity qualifications, as their identification relies on strict matching rules. An example of a SPARQL query for extracting direct qualifications is shown in Listing \ref{lst:sparql-direct}. The query binds a selected component's part number (\texttt{componentPn}) and checks for the existence of a \texttt{QUALIFICATION\_CARD} with the same \texttt{qualificationPn}, as well as identical \texttt{pkgCode}, \texttt{spkgCode}, and \texttt{manufacturerName}. Additional metadata, such as qualification status, type, and documentation, is also extracted using the optional clause ensuring resilience to missing values in the underlying data. The methodology for qualifications by similarity is equivalent but without binding the \texttt{qualificationPn} to the \texttt{componentPn}.

In contrast, alternative qualifications require a more flexible retrieval strategy, as they are not based on fixed attribute identity. Since the criteria for this type of qualification vary depending on the component type and are based on unstructured data (e.g., the description of the assembly process), one approach is to use symbolic queries to filter qualifications based on the available structured data and have the designer (i.e., a domain expert) manually review each candidate. This process, however, can further be optimized by leveraging an LLM's capability to process unstructured data and rank such candidates based on their relevance. A natural approach to implement this is through a vector-based retrieval mechanism. Both PLM-DB components and qualification entities are converted to JSON with their relevant attributes (e.g., assembly description, mounting type, manufacturer, package dimensions, and others) and are embedded into vector representations using \texttt{multilingual-e5-large-instruct}, an open-source language embedding model. Finally, given a PLM-DB component, a SPARQL query is first used to filter qualifications that are not compliant with the criteria and then the embeddings of the resulting qualification candidates are ranked by cosine similarity to the embedding of the component to be subsequently reviewed by the designer. This step is only done when no direct or by-similarity qualifications were found by their respective queries.

\begin{lstlisting}[language=SPARQL, caption={SPARQL query for retrieving directly qualified components and their respective qualification data for a given Part Number (PN).}, label={lst:sparql-direct}, abovecaptionskip=10pt]
PREFIX rdf: <http://www.w3.org/1999/02/22-rdf-syntax-ns#>  
PREFIX tasi: <http://tasi.com#>  

SELECT ?generic_pn ?package_nbr ?subpackage_nbr ?manufacturer_name 
       ?conf_coating ?conf_substrate ?conf_mounting 
       ?q_number ?q_description ?q_status ?q_type ?q_documents
WHERE {
  ?c a tasi:PLMDB_COMPONENT; tasi:componentPn "{selected_value}";
     tasi:pkgCode ?package_nbr; 
     tasi:spkgCode ?subpackage_nbr;
     tasi:manufacturerName ?manufacturer_name .
  ?qc a tasi:QUALIFICATION_CARD; tasi:qualificationPn "{selected_value}";
     tasi:pkgCode ?package_nbr; 
     tasi:spkgCode ?subpackage_nbr;
     tasi:manufacturerName ?manufacturer_name .
  OPTIONAL {
      ?qc tasi:qID ?q_number; tasi:status ?q_status; 
          tasi:documentation ?q_documents; 
          tasi:qualificationDesc ?q_description;
          tasi:qualificationType ?q_type;
          tasi:conformalCoating ?conf_coating; 
          tasi:substrateMaterial ?conf_substrate; 
          tasi:assemblyType ?conf_mounting .
      ?c tasi:componentGenPn ?generic_pn .
  } 
}
\end{lstlisting}

\section{Evaluation and Impact}
\label{impl}
This section presents a benchmarking experiment assessing the performance of a RAG-based system as an alternative to our pipeline. We then provide a cost-benefit analysis, highlighting both the strengths and limitations of each method across different scales of use.

\medskip\noindent\textbf{Comparison with a RAG-based approach. }
Given the increasing popularity of Retrieval-Augmented Generation (RAG) in enterprise settings~\cite{heredia2024,an2024}, it is natural to consider whether such an approach could serve as a viable alternative to our more structured pipeline. RAG combines pre-trained language models with a retrieval mechanism that fetches relevant information from a vector database at query time, appending them to the model’s prompt to enrich its answers with external knowledge~\cite{rag_openbook}. This allows LLMs to answer questions based on up-to-date and domain-specific data, even if that data was not seen during training. These types of systems offer a very quick deployment, making them especially appealing in industrial scenarios where timelines are constrained. This is because they can operate directly on heterogeneous and unstructured data and thus do not require schema alignment, data cleaning, or ontology engineering.

Based on this idea, we implemented a RAG pipeline tailored to the qualification retrieval task. Both the PLM-DB (containing electronic components) and QC (containing qualification records) were converted into JSON objects and embedded using \texttt{multilingual-e5-large-instruct}. Given a component from PLM-DB, the system ranks QC entries based on cosine similarity and selects the top 200 most similar records. These 200 entries are then added as context into a prompt for the LLM, which is asked to identify and classify qualification matches according to our predefined types. Specifically, for direct and by-similarity qualifications, we use the definition described in Section \ref{mot}, while for alternative qualifications, we consider as a rule that only the “Package Code” and “Manufacturer” must match.
\begin{table}[t]
    \caption{Performance of the RAG-based approach with different LLMs for direct, by-similarity, and alternative qualifications.}
    \label{table:metrics}
    \medskip
    \centering
    \begin{tabularx}{\linewidth}{l l *{4}{>{\centering\arraybackslash}X}}
        \toprule
        \textbf{Model} & \textbf{Qual. Type} & \textbf{Precision} & \textbf{Recall} & \textbf{F1-score} & \textbf{IoU} \\
        \midrule
        \multirow{4}{*}{\textit{GPT-OSS-120B }} 
            & Direct      & 0.947 & 0.896 & 0.921 & 0.959 \\
            & Similarity  & 0.969 & 0.907 & 0.937 & 0.907 \\
            & Alternative & 0.516 & 0.773 & 0.619 & 0.596 \\
            & \textbf{Overall} 
                          & 0.866 & 0.886 & 0.876 & 0.846 \\
        \midrule
        \multirow{4}{*}{\textit{Mistral 3 24B}} 
            & Direct      & 0.055 & 0.654 & 0.101 & 0.287 \\
            & Similarity  & 0.152 & 0.306 & 0.203 & 0.143 \\
            & Alternative & 0.024 & 0.384 & 0.046 & 0.131 \\
            & \textbf{Overall} 
                          & 0.081 & 0.335 & 0.131 & 0.237 \\
        \bottomrule
    \end{tabularx}
\end{table}
To evaluate the performance of the RAG approach, we conducted a benchmarking experiment using the results of our deployed VKG+LLM pipeline as ground truth. We selected a representative subset of 675 components from the PLM-DB and retrieved their qualification matches using both methods. In this subset, 17.48\% of components have never been qualified, while the remaining components have on average 0.63 direct qualifications, 7.98 by similarity, and 2.23 alternative qualifications. The system's performance was measured using standard metrics such as Precision, Recall, F1-score, and Intersection over Union (IoU). The results using two different open-source LLMs are reported in Table~\ref{table:metrics}. In general, RAG using the GPT model achieves some promising results in our use case, particularly for direct and by-similarity qualifications, with F1-scores of 92.07\% and 93.71\%, respectively. However, for alternative qualifications, the performance is noticeably lower, with an F1-score of 61.88\% and IoU of 59.64\%. The variant based on Mistral 3 24B performs substantially worse, with an overall F1-score of 13.06\%.
This could be due to the model's difficulty in handling a large context window with hundreds of qualifications, which is a challenging task for smaller LLMs. While performing a model call for each component-qualification pair might improve accuracy, we did not consider this possibility as it is computationally infeasible, taking more than one hour per component.

Regarding the applicability of this approach in a real industrial scenario, it is evident that while the RAG-based method can serve as a suggestion tool for designers, it still necessitates a human-in-the-loop component for manual verification. A possible approach would be to provide designers with the 200 most similar entries to manually review and verify the qualifications suggested. This would still be an improvement compared to manual search since we found that 78.5\% of qualifications for a component are found within the top 50 results, 93.4\% within the top 100, and 99.8\% within the top 200. However, whether this is the best long-term strategy from a business perspective is discussed in the subsequent paragraph.

\begin{figure}[t!]
    \centering
    \includegraphics[width=1\textwidth]{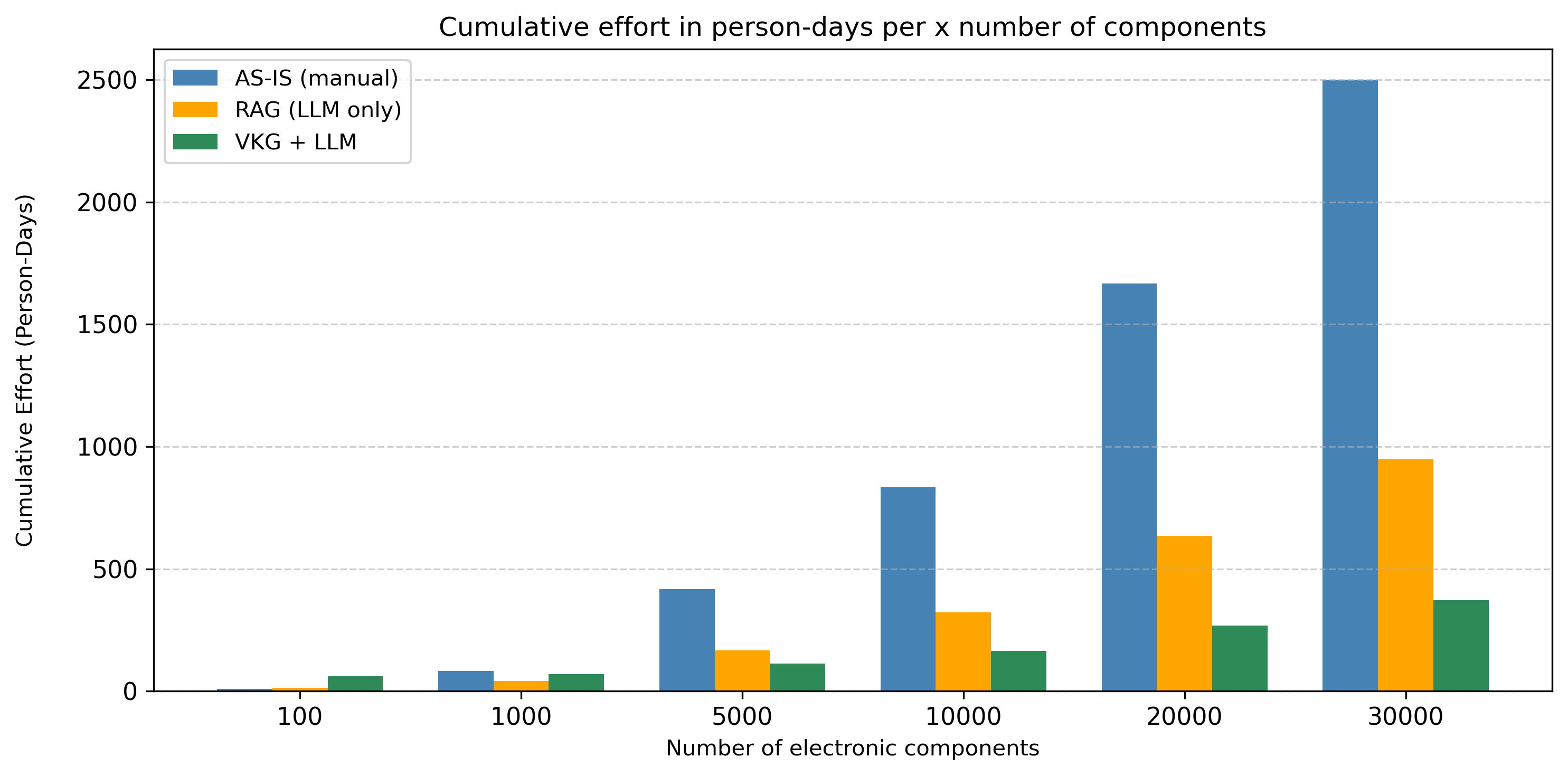}
    \hfill
    \includegraphics[width=1\textwidth]{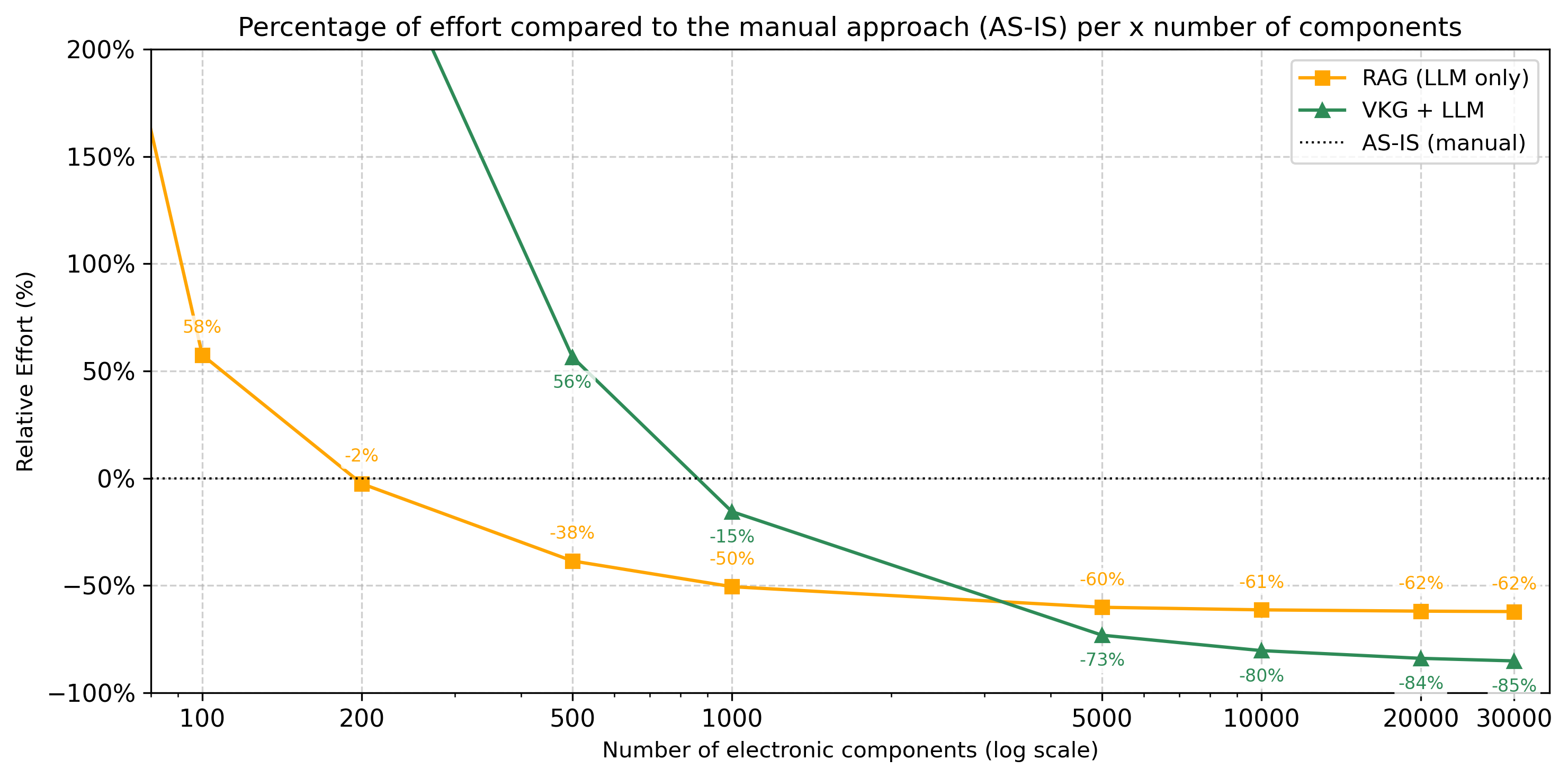}
    \caption{Cumulative effort in person-days required for each approach as a function of the number of electronic components, considering system setup time and human-in-the-loop validation. The top figure
    shows the absolute effort of the three pipelines, while the bottom figure shows the relative effort compared to the manual (AS-IS) baseline. The VKG+LLM pipeline requires higher setup time but becomes significantly more efficient at scale, reducing effort by over 70\% beyond 5000 components.}
    \label{fig:cost-model-1-template}
\end{figure}

\medskip\noindent\textbf{Cost-Benefit Analysis. }
While the RAG-based approach requires minimal engineering effort and was implemented in just a few days, our structured pipeline involved several weeks of collaboration with domain experts for data cleaning, ontology modeling, and implementation. To quantify the trade-offs between approaches, we developed a cost model that estimates the effort required to extract qualification manually, using RAG with the GPT model, or our VKG+LLM pipeline.
For the manual procedure (AS-IS), there are no upfront costs, but each PLM-DB component requires an average of 40 person-minutes to manually review qualifications and check for compliance. RAG took only about 10 person-days to design and implement, but it still relies on humans to validate LLM answers and revert to manually search within ranked qualifications in case these answers are incorrect. This reduces the average effort to roughly 15 person-minutes per component.
Regarding the VKG+LLM approach, it took approximately 60 person-days to set up, but direct and by-similarity qualifications are automated, and a manual review is required only in case of alternative qualifications, resulting in an overall average of about 5 person-minutes per component. 

The plots in Figure \ref{fig:cost-model-1-template} show the absolute and relative effort (y-axis) over an increasing amount of components (x-axis). The AS-IS approach is advantageous for only up to 200 components, after which the RAG approach becomes the most convenient, offering a decent trade-off between setup and validation effort. Still, it is eventually overtaken by the VKG+LLM pipeline, which, despite its higher upfront time investment, reduces cumulative effort significantly, offering over 80\% savings at the scale of 10000 components.

While the VKG+LLM approach is more efficient at higher scales (i.e., in the long term), it is also interesting to see that LLMs alone were able to provide a good effort reduction even with a small upfront investment. This emphasizes their usefulness for rapid prototyping (i.e., developing proofs-of-concept for large-scale solutions). 
As the performance of LLMs continues to improve, we expect their break-even point with respect to structured approaches like VKG+LLM to shift further toward higher scales. However, the increased computational requirement would introduce new cost dimensions such as hardware and energy consumption, which for simplicity are not considered in this analysis.

\section{Uptake and Lessons Learned}
\label{lessons}
This section describes our deployment experience at Thales Alenia Space and discusses the broader organizational and strategic implications. We then reflect on the main lessons learned working on this case study.

\medskip\noindent\textbf{Deployment Details. }
The tool was successfully deployed in a production-ready state within an operational aerospace industrial environment, representing a significant milestone in the transition from research to practice. The system is currently in active use, with authenticated access granted to approximately 50 users across different departments such as AT (Advanced Technologies), M\&P (Materials and Processes), IPE (Industrial Product Engineering), DESIGN, and ENG (Engineering). These groups represent key functional areas involved in component qualification, evaluation, and integration. Before deployment, a structured training program was conducted that involved more than 50 individuals from the same departments. The training aimed to familiarize users with the application interface and clarify its role in supporting the identification and comparison of qualification cases across different component families.

This training phase was instrumental in fostering an early understanding of the system's capabilities and aligning expectations regarding its use. Currently, the application is in adoption and under continuous evaluation to assess whether it can be integrated into existing workflows and decision-making processes.

A broader European-scale deployment is being considered and a feasibility study is being discussed to explore this possibility. The study should address several critical dimensions, including data source heterogeneity across institutions, interoperability with existing repositories, access control models for inter-organizational usage, and the technical infrastructure needed to support secure and scalable deployment across multiple industrial actors.

Furthermore, the deployment has initiated strategic conversations with stakeholders at the European level, including the European Space Agency (ESA), regarding the potential development of a shared service to consolidate qualification knowledge. This initiative aims to facilitate the reuse of qualification experiences across the space industry, opening pathways toward greater cross-organizational collaboration and mutualization of noncompetitive technical assets.

\medskip\noindent\textbf{Lessons Learned. }
Several key lessons emerged from the experience presented in this paper. In real-world KG implementation, data cleaning proved to be as crucial and even more time-consuming than data modeling. Without thorough data cleaning, even a well-designed semantic model cannot support effective querying due to underlying data inconsistency. Leveraging LLMs significantly streamlined this process, substantially reducing manual efforts in correcting heterogeneities and enabling knowledge engineers to focus primarily on modeling tasks. 
Furthermore, another interesting insight was the effectiveness of integrating symbolic queries with embedding-based methods. This allowed the system to maximally exploit both structured and unstructured data and address complex retrieval scenarios. Collaboration with domain experts and iterative development also proved to be essential for validating the ontology as well as fine-tuning the LLM prompts used in the data cleaning phase, ensuring that the LLM answers align with the desired outcome.

Another important lesson was that RAGs can be very valuable for rapidly testing proofs of concept, given the speed and ease with which these systems can be implemented. Still, they are less suitable for long-term use, especially in critical domains such as aerospace, where accuracy is paramount and errors carry high costs. In such scenarios, the initial advantage provided by quick setup is offset by the continual need for human oversight and real-time correction of system outputs. On the other hand, semantic technologies have been shown to be efficient and scalable to tens of thousands of components which corresponds to the scale of our retrieval task. 

Finally, many of these lessons are not limited to the electronic components domain. The benefits of semi-automated data cleaning, integration of symbolic and vector-based methods and using RAGs for rapid prototyping are all potentially applicable across other industrial qualification processes and data integration challenges in general. For instance, within Thales Alenia Space, we plan to implement a similar methodology for mechanical qualifications, which exhibit similar issues related to fragmented and unstructured data.

\section{Conclusion and Future Work}
\label{concl}
This paper presents a novel semantic data integration pipeline that enhances the VKG paradigm through the use of LLMs for semi-automated data cleaning and to enable the support of vector-based search. The pipeline was developed to address a concrete aerospace use case involving the retrieval of electronic component qualifications from heterogeneous data silos. It is currently deployed and actively used at Thales Alenia Space, demonstrating considerable operational benefits, and is under evaluation for broader adoption at the European level. Comparative analysis also highlighted that the proposed approach is more efficient and scalable than purely LLM-based methods, such as RAG, which may be better suited for proofs of concept or short-term solutions.

In future work, we plan to explore the applicability of hybrid workflows in which LLMs also support knowledge engineering tasks such as ontology design and mapping generation. While this remains an active area of research, it is gaining increasing attention from the scientific community \cite{fathallah2024neon,zhang2024ontochat,SHIMIZU2025100862,VALCALVO2025104042,Laurenzi_Mathys_Martin_2024} and may offer a promising direction to mitigate the main weakness of the VKG+LLM framework, which was its high initial setup effort compared to a RAG baseline.

\paragraph*{Supplemental Material Statement:} The code used in some of the experiments, including all the prompts used for the LLMs, is made available to enhance reproducibility and potential reuse for future research. The data itself and the implementation of the deployed tool cannot be shared as they are part of confidential material belonging to Thales Alenia Space. The available resources can be accessed in our GitHub Repository at \url{https://github.com/Antonio-Dee/tasi-qual}.

\small\medskip\noindent\textbf{Acknowledgments.} Antonio De Santis's doctoral scholarship is funded by the Italian Ministry of University and Research (MUR) under the National Recovery and Resilience Plan (NRRP), by Thales Alenia Space, and by the European Union (EU) under the NextGenerationEU project. Conference attendance was partially supported through a Young Researcher and Innovator Conference Grant under COST Action CA23147 GOBLIN – Global Network on Large-Scale, Cross-domain and Multilingual Open Knowledge Graphs, supported by COST (European Cooperation in Science and Technology, https://www.cost.eu).

%
%
%
\bibliographystyle{splncs04}
\bibliography{tasi}
\end{document}